\documentclass[aps,longbibliography,showpacs,twocolumn,superscriptaddress,amsmath,amssymb,verbatim]{revtex4-1}

\usepackage[thinlines]{easytable}
\usepackage{graphicx}
\usepackage{lmodern}
\usepackage{epstopdf}
\usepackage{dcolumn}
\usepackage{bm}
\usepackage{subfigure}
\usepackage{amsmath} 
\usepackage[pdftex,colorlinks=true,citecolor=blue,linkcolor=blue,urlcolor=blue,bookmarks=true]{hyperref}

\usepackage{makecell}

\usepackage{color}
\usepackage{textcomp}
\definecolor{red}{rgb}{1,0,0}

\definecolor{blue}{rgb}{0,0,1}

\definecolor{green}{rgb}{0,1,0}

\begin{document}
	\preprint{APS}

\title{ Magnetic properties  and spin dynamics in a spin-orbit driven $J_{\rm eff}$= 1/2 triangular lattice antiferromagnet}

\author{J. Khatua}
\affiliation{Department of Physics, Indian Institute of Technology Madras, Chennai 600036, India}
\author{S. Bhattacharya}
\affiliation{Universit\'e Paris-Saclay, CNRS, Laboratoire de Physique des Solides, 91405, Orsay, France}
	\author{A. M. Strydom}
\affiliation{Highly Correlated Matter Research Group, Department of Physics, University of Johannesburg, PO Box 524, Auckland Park 2006, South Africa}
\author{A. Zorko}
\affiliation{Jo\v{z}ef Stefan Institute, Jamova c.~39, SI-1000 Ljubljana, Slovenia}
\affiliation{Faculty of Mathematics and Physics, University of Ljubljana, Jadranska u.~19, SI-1000 Ljubljana, Slovenia}
\author{J. S. Lord}
\affiliation{ISIS Facility, STFC Rutherford Appleton Laboratory, Didcot OX11 0QX, UK}
\author{A. Ozarowski}
\affiliation{National High Magnetic Field Laboratory, Florida State University, Tallahassee, Florida 32310, USA}
\author{E. Kermarrec}
\affiliation{Universit\'e Paris-Saclay, CNRS, Laboratoire de Physique des Solides, 91405, Orsay, France}
\author{P. Khuntia}
\email[]{pkhuntia@iitm.ac.in}
\affiliation{Department of Physics, Indian Institute of Technology Madras, Chennai 600036, India}
\affiliation{Quantum Centre of Excellence for Diamond and Emergent Materials, Indian Institute of Technology Madras,
	Chennai 600036, India.}
\date{\today}

\begin{abstract}
	Frustration induced strong quantum fluctuations accompanied by spin-orbit coupling and crystal electric field  can give rise to rich and  diverse magnetic phenomena associated with unconventional low-energy excitations  in rare-earth based quantum magnets.
 Herein, we present crystal structure, magnetic susceptibility, specific heat, muon spin relaxation ($\mu$SR), and electron spin resonance (ESR)  studies on the  polycrystalline samples of Ba$_{6}$Yb$_{2}$Ti$_{4}$O$_{17}$  in which Yb$^{3+}$ ions constitute a perfect triangular lattice in $ab$-plane without detectable  anti-site disorder between atomic sites. 
 The Curie-Weiss fit of low-temperature magnetic susceptibility data suggest the spin-orbit entangled $J_{\rm eff}$ = 1/2 degrees of freedom of Yb$^{3+}$ spin  with weak antiferromagnetic exchange interactions in the Kramers doublet ground state. The zero-field specific heat data reveal the presence of long-range magnetic order at $T_{N}$ = 77 mK which is suppressed in a magnetic field $\mu_{0}H \geq 1$ T.
The broad maximum in specific heat is attributed to the Schottky anomaly implying the Zeeman splitting of the Kramers doublet ground state.  The ESR measurements suggest the presence of anisotropic exchange interaction between the moments of Yb$^{3+}$ spins and the well separated Kramers doublet state. $\mu$SR experiments reveal a fluctuating state of Yb$^{3+}$ spins in the temperature range 0.1 K $\leq$ $T$ $\leq$ 10 K owing to depopulation of crystal electric field levels, which  suggests that the Kramers doublets are well separated consistent with thermodynamic and ESR results. In addition to the intraplane nearest-neighbor superexchange interaction, the interplane exchange interaction and anisotropy are expected to stabilize long-range ordered state in this triangular lattice antiferromagnet.  
\end{abstract}
\maketitle
\section{Introduction}
Quantum materials wherein the interplay between competing degrees of freedom, frustration-induced strong quantum fluctuations, and quantum entanglement is prominent have drawn considerable attention due to their connection to the emergence of exotic quantum phenomena with potential to address some of the recurring themes in quantum condensed matter \cite{Balents2010,doi:10.1146/annurev-conmatphys-031218-013423,Keimer2017,Witczak-Krempa2014,Glasbrenner2015,KHATUA20231}. One noteworthy example is the quantum spin liquid (QSL), in which strongly entangled electron spins do not exhibit long-range magnetic order down to absolute zero temperature despite strong spin correlations \cite{Savary_2016,Broholmeaay0668}. Originally,
the QSL state was proposed by Anderson for a system composed of $S$ = 1/2 spins on a two-dimensional triangular lattice antiferromagnet \cite{ANDERSON1973153}. Its materialization has been proposed in a few triangular lattice antiferromagnets with next-nearest interaction, and anisotropic exchange interaction \cite{Klanjsek2017,Arh2022,PhysRevB.93.140408,PhysRevLett.91.107001,PhysRevB.74.014408,PhysRevX.9.021017}.
Furthermore, the quantum nature of this entangled state offers an  outstanding track to study gauge theories incorporating fractionalized excitations, as well as its relevance in the field of quantum computing \cite{RevModPhys.89.025003,RevModPhys.80.1083,KITAEV20062}. Despite the fact that several QSL candidate materials show analogous behavior in various observables \cite{KHUNTIA2019165435,Han2012,PhysRevLett.113.247601,PhysRevLett.116.107203,Gao2019}, understanding the microscopic spin Hamiltonian to identify fractionalized excitations remains a significant challenge because of the presence of extra exchange couplings, intrinsic disorder, and unavoidable defects in real materials \cite{PhysRevLett.118.087203,PhysRevB.93.140408,Khatua2022,RevModPhys.80.1083,PhysRevMaterials.5.034419}.\\
  In this context, rare-earth-based materials featuring $J_{\rm eff}$ = 1/2 quantum spins decorated on geometrically frustrated lattices offer a promising avenue for the experimental realization of exotic quantum phenomena \cite{https://doi.org/10.1002/qute.201900089,Arh2022,PhysRevB.106.104404,PhysRevB.106.104408,PhysRevX.12.021015,Gao2019,PhysRevLett.120.207203}. Moreover, it has been suggested that anisotropic magnetic interactions between $J_{\rm eff}$ = 1/2 moments in  the Kramers crystal-field ground state of these magnetic materials  play an important role in the materialization of QSL state, in contrast to isotropic interactions between pure $S$ = 1/2 moments on the triangular lattice Heisenberg model, which tend to promote a 120$^\circ$ long-range magnetic ordered state \cite{PhysRevB.50.10048}.  For example, the rare-earth triangular antiferromagnets YbMgGaO$_{4}$ \cite{Li2015,PhysRevLett.117.097201,Shen2016} and chalcogenides NaYb$Ch_{2}$ ($Ch$ = O, S, Se )  \cite{WeiweiLiu117501,PhysRevB.100.220407,PhysRevB.98.220409,PhysRevB.100.224417,Bordelon2019,PhysRevB.100.144432,PhysRevB.98.220409,PhysRevB.100.241116,PhysRevX.11.021044} with Yb$^{3+}$ ions with $J_{\rm eff}$ = 1/2 moment have attracted considerable interest due to the emergence of a QSL ground state brought about by the presence of dominant easy-plane anisotropic magnetic interactions and spin frustration \cite{PhysRevLett.120.087201,PhysRevX.8.031028,PhysRevLett.120.087201}. In contrast to QSL state with dominant easy-plane anisotropic magnetic interactions, the spin-liquid candidates on a perfect triangular lattice with dominant easy-axis anisotropy are very rare \cite{PhysRev.79.357} and have been found only very recently in NdTa$_{7}$O$_{19}$ which appears to be the first realization of spin-liquid ground state with Ising-like magnetic correlations driven by strong easy-axis anisotropic magnetic interaction \cite{Arh2022,ulaga2023finitetemperature}.
 Furthermore, the impact of dipolar magnetic interactions to the observed spin excitation continuum, widely considered as the most robust evidence of spin fractionalization in insulators, remains a dynamic and actively researched domain in quantum condensed matter \cite{Yao,doi:10.1146/annurev-conmatphys-031016-025218}.\\ In addition to the anisotropic exchange interactions, there have been proposals for a dominant dipolar interaction-driven spin-liquid state in certain triangular lattice antiferromagnets, such as Yb(BaBO$_{3}$)$_{3}$ \cite{PhysRevB.106.014409,PhysRevB.102.045149,PhysRevB.104.L220403} and $A$BaYb(BO$_{3}$)$_{2}$ (where A = Na, K) \cite{PhysRevMaterials.3.094404,PhysRevB.103.104412}. In addition to their dynamic ground states, the phase transitions observed in triangular lattice antiferromagnets are of significant interest for gaining insights into intriguing quantum phenomena such as Berezinskii-Kosterlitz-Thouless (BKT) physics \cite{J1973,Gao2022}.
 For example, the observed quasi long-range magnetic ordered state in the two-dimensional triangular lattice material TmMgGaO$_{4}$ may represent the manifestation of the BKT phase \cite{Hu2020,Li2020,PhysRevB.103.064424}. Recently, it has been suggested that the dipolar spin liquid candidate KBaGd(BO$_{3}$)$_{2}$ is a promising candidate to  host a BKT phase and an unconventional quantum critical point \cite{xiang2023dipolar}. Furthermore, the rare-earth-based magnets with weakly coupled magnetic moments, approaching the paramagnetic limit, have shown potential for lowering system temperatures through adiabatic demagnetization cooling, as observed in triangular lattice antiferromagnets such as KBaGd(BO$_{3}$)$_{2}$ \cite{PhysRevB.107.104402} and KBaYb(BO$_{3}$)$_{2}$ \cite{Tokiwa2021}.
 Interestingly, in such rare-earth quantum magnets, anisotropic magnetic interaction induced by spin-orbit coupling, crystal-electric field  of  localized 4\textit{f} electrons and lattice symmetry offer a viable platform to host unconventional phases including spiral spin liquid \cite{PhysRevLett.100.136402,PhysRevLett.130.166703}. In a similar vein,  Kitaev spin liquid state induced by bond-dependent anisotropic  magnetic interactions \cite{Razpopov2023,Kim2023,PhysRevB.104.L100420}, and multipolar orders \cite{PhysRevB.94.201114} in spin-orbit entangled $J_{\rm eff}$ = 1/2 quantum magnets remains largely unexplored.    The current effort is devoted towards the design,  discovery and investigation  of two- and three dimensional spin-orbit entangled 4$f$ based frustrated magnets potential to host myriads of exotic  quantum phases that may aid for the establishment of  new theoretical paradigms in quantum condensed matter. In order to realize rich variety of spin-orbit driven quantum phenomena as a result of competing exchange interaction, crystal electric field and anisotropic interaction, it is therefore essential in exploring disorder free rare-earth based quantum magnets. 
\\
Herein, we report the crystal structure and magnetic properties of a unexplored triangular lattice antiferromagnet Ba$_{6}$Yb$_{2}$Ti$_{4}$O$_{17}$, (henceforth, BYTO) which crystallizes in the hexagonal space group $P6_{3}/mmc$ where the Yb$^{3+}$ ions constitute a perfect triangular lattice in the $ab$-plane. The Rietveld analysis of x-ray diffraction data shows the absence of anti-site disorder between the crystallographic sites. Magnetic susceptibility data suggest that Yb$^{3+}$ spins acquire $J_{\rm eff}$ = 1/2 degrees of freedom and the presence of a weak antiferromagnetic exchange interaction between rare-earth moments. The zero-field specific heat data show an anomaly possibly due to the long-range magnetic order at $T_{N} = 77 $ mK which disappears in a magnetic field $\mu_{0}H \geq 1 $ T.
 $\mu$SR experiments suggest the lack of long-range magnetic order down to 80 mK, while spin fluctuations are observed up to 10 K.
 The anisotropic magnetic exchange interaction between $J_{\rm eff}$ = 1/2 moments of Yb$^{3+}$ spins was suggested from  electron spin resonance measurements. 
The combination of magnetic anisotropy, intraplane, and interplane exchange interactions are expected to stabilize a long-range magnetic ordered state in this antiferromagnet.  
\begin{figure*}
	\centering
	\includegraphics[width=\textwidth]{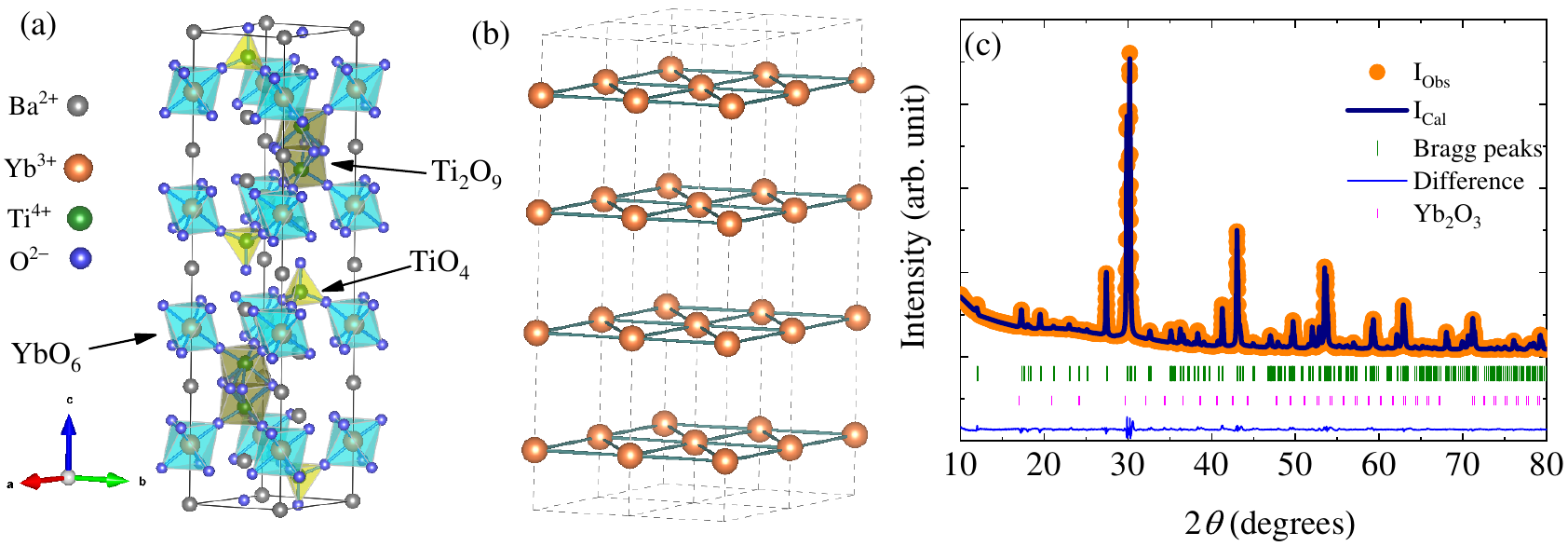}
	\caption{(a) Schematics of the unit cell of Ba$_{6}$Yb$_{2}$Ti$_{4}$O$_{17}$. The oxygen atoms coordinate with Yb$^{3+}$ ions in an octahedral arrangement (blue). Two top and  bottom layers of YbO$_{6}$ octahedra are connected through Ti$_{2}$O$_{9}$ double octahedra while two middle layers of YbO$_{6}$ connected through TiO$_{4}$ tetrahedra. (b) Two-dimensional triangular layers of Yb$^{3+}$ ions  arranged in the unit cells.   (c) The Rietveld refinement of powder x-ray diffraction data recorded at room temperature. The experimentally observed points (scattered orange points), calculated Rietveld refinement profile (solid navy line), Bragg reflection positions of BYTO (olive vertical bars), Bragg reflection position of Yb$_{2}$O$_{3}$ (pink bars) and the difference between observed and calculated intensity (solid blue line) are shown.}{\label{xrd}}.
\end{figure*}
\\
\section{Experimental details}  Polycrystalline samples of BYTO were prepared by a conventional solid state  method. Prior to use, we preheated a stoichiometric amount of  BaCO$_{3}$ (Alfa Aesar, 99.997 \text{\%}), and Yb$_{2}$O$_{3}$ (Alfa Aesar, 99.998 \text{\%}) at 200$^\circ$C and 800$^\circ$C, respectively   to prevent moisture contamination. The stoichiometric mixtures were  pelletized and sintered at 1400$^\circ$C for 72 hours with intermittent grindings.  The phase purity  of the final product was confirmed by the x-ray diffraction at 300 K by employing Rigaku smartLAB x-ray diffractometer with Cu K$\alpha$ radiation ($\lambda $ = 1.54 {\AA}). Magnetization measurements were carried out using a Quantum Design, SQUID (MPMS) in the temperature range 2 K $\leq$ \textit{T} $\leq$ 350 K in several magnetic fields.  Specific heat measurements were performed  using a Quantum Design, physical properties measurement system (PPMS) in the temperature range 2 K $\leq$ \textit{T} $\leq$ 270 K and in magnetic fields up to 7 T. In addition, specific heat measurements were carried out in the temperature range
0.049 K $\leq$ \textit{T} $\leq$ 4 K in 0 T, 1 T, 3 T in  a dilution refrigerator using a Dynacool PPMS instrument from Quantum Design,
San Diego, USA. Thermal conductivity measurements were conducted in the temperature range 2 K $\leq$ \textit{T} $\leq$ 300 K in magnetic fields up to 7 T by two probe method also using Quantum Design, PPMS. \\
\begin{table}
	\caption{\label{tab:table1 } Refined structural parameters based on x-ray diffraction data at 300 K. (space group: $P$6$_{3}$/$mmc$, $ \alpha$ = $ \beta$ = 90.0$ ^{0}$, $\gamma$ = 120.0$ ^{0}$), $a$ = $b$ = 5.907 {\AA}, $c$ = 29.426 {\AA}
		and $\chi^{2}$ = 4.22, R$_{\rm wp}$ = 5.88, R$_{\rm p}$ = 3.39, and R$ _{\rm exp}$ = 2.86)}
	
	\begin{tabular}{c c c c c  c c} 
		\hline \hline
		Atom & Wyckoff position & \textit{x} & \textit{y} &\textit{ z}& Occ.\\
		\hline 
		Yb & 4e & 0 & 0 & 0.126 & 1 \\
		Ba$_{1}$ & 2a & 0 & 0 & 0 & 1 \\
		Ba$_{2}$ & 24l & 0.666& 0.333 & 0.089 & 1 \\
		Ba$_{3}$ & 24l & 0.333& 0.666 & 0.182 & 1 \\
		Ba$_{4}$ & 2b & 0& 0 & 0.25 & 1 \\
		Ti$_{1}$ & 12k & 0.666& 0.333 & $-$0.055 & 1 \\
		Ti$_{2}$ & 12k & 0.666& 0.333 & 0.203 & 1 \\
		O$_{1}$ & 12k & 0.666& 0.333 & $-$0.006 & 1 \\
		O$_{2}$ & 12k & $-$0.347& $-$0.173 & 0.077 & 1 \\
		O$_{3}$ & 12k & 0.346& 0.173 & $-$0.170 & 1 \\
		O$_{4}$  & 12j &0.573 &0.014 &0.25 &1 \\
		
		\hline
	\end{tabular}
\end{table}
$\mu$SR experiments in zero field and in longitudinal magnetic fields were performed on the MuSR spectrometer at the ISIS pulsed neutron and muon source at Rutherford Appleton Laboratory, UK. The powder sample ($\sim$ 1 gm) was mixed with a small amount of GE-varnish and fixed to a silver plate. A 25 $\mu$m-thick silver foil was placed on top of the sample to maximize the number of implanted muons in the sample and to reduce the thermal radiation. A dilution fridge was used for the temperature range 0.08 K to 4~K, and subsequently an helium flow cryostat to reach temperatures from 1.4~K to 300~K.
Electron spin resonance (ESR) measurements were performed at the National High Magnetic Field Laboratory, Tallahassee, USA on a polycrystalline sample using a custom-made transmission-type ESR spectrometer with homodyne detection.
The measurements were conducted in the Faraday configuration at the irradiation frequency of 212\,GHz.
The magnetic field was swept between 2 and 10\,T using a superconducting magnet and the temperature was varied between 5 and 200\,K using a continuous-flow He cryostat.
A standard field-modulation technique was used with the modulation field of about 2\,mT.
\section{results}
\subsection{Rietveld refinement  and crystal structure}To examine the phase purity and crystal structure, we carried out the Rietveld refinement of room temperature x-ray diffraction data using GSAS software \cite{doi:10.1107/S0021889801002242}. The XRD results suggest that the polycrystalline BYTO samples contain a tiny fraction of unreacted magnetic Yb$_{2}$O$_{3}$ impurity, which has a minimal influence on the overall magnetic properties of the material under examination.
A similar situation with the unavoidable Yb$_{2}$O$_{3}$ secondary phase in polycrystalline samples has been observed in a few other Yb-based magnets \cite{PhysRevB.107.224416,PhysRevB.106.075132,PhysRevB.108.054442}. To accurately determine the percentages of the main and secondary phase, a two-phase  Rietveld refinement was performed. The crystallographic parameters of isostructural compound Ba$_{6}$Y$_{2}$Ti$_{4}$O$_{17}$ were incorporated as a reference to perform the Rietveld refinement for the main phase \cite{Kuang2002}. 
Figure~\ref{xrd} (c) depicts the Rietveld refinement pattern of the x-ray diffraction data indicating that our polycrystalline samples consist of 98.5 \text{\%} BYTO and 1.5 \text{\%} Yb$_{2}$O$_{3}$ phases.
The Rietveld refinement results indicate the compound BYTO crystallizes in the 12H hexagonal structure with space group $P$6$_{3}$/$mmc$ without inter-site mixing between constituents ions. The obtained lattice parameters are  $a$ = $b$ = 5.907 {\AA}, $c$ = 29.426 {\AA}, $\alpha$ = $\beta$ = 90$^\circ$  and $\gamma$ = 120$^\circ$. The estimated fractional  atomic  coordinates and  \textit{R}  factors  are summarized   in table \ref{tab:table1 }.\\ Figure \ref{xrd} (a) depicts the refined crystal structure of BYTO drawn using VESTA software \cite{https://doi.org/10.1107/S0021889808012016}. The  nearest neighbor magnetic Yb$^{3+}$  ion (Yb-Yb $\approx$ 5.907 {\AA}) constitutes two  dimensional triangular layers  stacked along the \textit{c}-axis (see Figure~\ref{xrd} (b)) similar to the extensively studied triangular lattice antiferromagnet YbMgGaO$_{4}$ with Yb-Yb distance of approximately  3.4 {\AA}  \cite{Li2015}. As shown in Figure~\ref{xrd} (b), one unit cell of BYTO is composed of four triangular layers of Yb$^{3+}$ ions where the first and last two  layers from the bottom are separated  by the inter-planar distance of 7.26 {\AA} while the middle two layers are separated by the inter-planar distance 7.45 {\AA}.  In BYTO, the Yb$^{3+}$ ion occupies a single crystallographic site (4$e$) of the hexagonal lattice and forms a distorted YbO$_{6}$ tetrahedron (see Figure~\ref{xrd} (a)) with local O$^{2-}$ ions. Due to such an octahedral environment, Yb$^{3+}$ ions are exposed to a strong crystal electric field, which splits the degenerate multiplet of $J$ = 7/2 into four Kramers doublet states of Yb$^{3+}$ ions with spin-orbit entangled $J_{\rm eff}$ = 1/2 moment.
\begin{figure*}
	\centering
	\includegraphics[width=\textwidth]{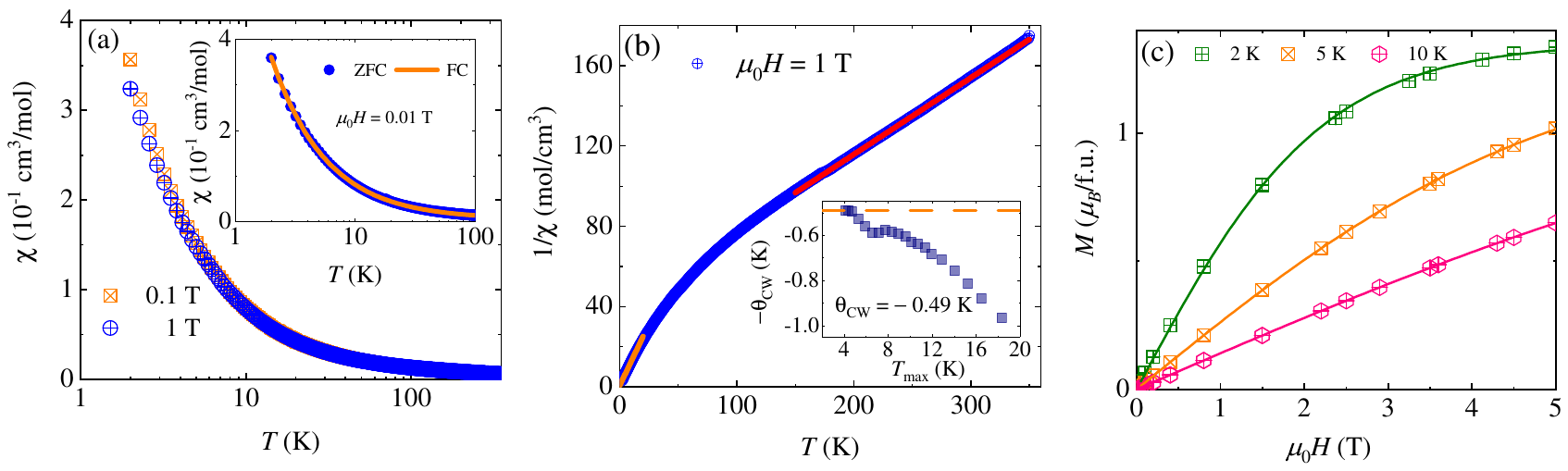}
	\caption{(a) The temperature dependence of magnetic susceptibility $\chi(T)$ in two different magnetic fields. The inset shows the temperature dependence of zero-field cooled (ZFC) and field-cooled (FC)  magnetic susceptibility measured in $\mu_{0}H$ = 100 Oe. (b) The temperature dependence of inverse  magnetic susceptibility (1/$\chi(T)$). The red and orange lines are the Curie-Weiss fit to the  high-temperature and low-temperature inverse magnetic susceptibility data, respectively. 
		The inset shows the estimated Curie-Weiss temperature obtained by varying upper limit of temperature range while the constant value of  Curie-Weiss temperature at low-temperature is shown by dotted orange line. 
		(c) Magnetization as a function of external magnetic field at several temperatures. The solid lines are the Brillouin function fit for paramagnetic Yb$^{3+}$ spins with $J_{\rm eff}$ = 1/2.  }{\label{BYTO2}}.
\end{figure*}
\\In BYTO, the intra-plane YbO$_{6}$ octahedra are not interconnected through a common O$^{2-}$ ions, as observed in other triangular lattice antiferromagnets such as YbMgGaO$_{4}$ or NaYb$X_{2}$ ($X$ = O, S, Se). Instead, they are linked by TiO$_{4}$ tetrahedra. On the other hand, the inter-plane YbO$_{6}$ octahedra are connected through Ti$_{2}$O$_{9}$ dimers, as illustrated in Figure \ref{xrd} (a). The larger unit cell of BYTO compared to YbMgGaO$_{4}$ is the primary reason for the formation of isolated YbO$_{6}$ octahedra, which could potentially reduce the strength of intra-plane magnetic exchange interactions between the Yb$^{3+}$ moments due to the increased bond length for the $f$-$p$-$d$-$p$-$f$ (Yb-O-Ti-O-Yb) hybridization. \\ The presence of weak exchange interactions, resulting from similar superexchange pathways, has been observed in several triangular lattice antiferromagnets, including NaBaYb(BO$_{3}$)$_{2}$ \cite{PhysRevMaterials.3.094404}, Ba$_{3}$Yb(BO$_{3}$)$_{3}$ \cite{PhysRevB.104.L220403,PhysRevB.103.104412}, and Ba$_{3}$YbB$_{9}$O$_{18}$ \cite{PhysRevB.106.104408} (as shown in Table \ref{tableIN}). However, this scenario differs somewhat for $3d$ transition ion-based magnets, as the electrons of 3$d$ ions are not highly localized in real space, unlike the 4$f$ ions. For instance, relatively strong exchange interaction  between Co$^{2+}$ ions  is observed in the 6H-hexagonal spin lattice of Ba$_{3}$CoSb$_{2}$O$_{9}$, despite the absence of a common oxygen mediator for intra-plane superexchange interactions between the $J_{\rm eff}$ = 1/2 moment of Co$^{2+}$ ions \cite{PhysRevLett.109.267206,Kamiya2018}.
From a structural perspective, it is worth noting that BYTO and Ba$_{3}$CoSb$_{2}$O$_{9}$ share several similarities, aside from differences in their stacking layers \cite{PhysRevLett.109.267206}. They have similar exchange paths, nearest-neighbor distances of 5.90 Å, and interlayer separations of 7.2 Å. However, the magnetic properties are anticipated to be distinct due to the localized nature and distinct anisotropy of Yb$^{3+}$ ions in BYTO.
\begin{table}[htb!]
	\centering
	\caption{Some promising rare-earth based frustrated triangular  lattice antiferromagnets and their  ground-state magnetic properties.}.\\
	\begin{tabular}{ | c | c | c | c | c |c|c|c|c|}
		\hline
		$\makecell{\rm 	Materials \\ (\rm symmetry)}$ & $\makecell{\rm Exchange \\ \rm path}$ &	$\makecell{\theta_{\rm CW} (\rm K)\\(\textnormal{ \rm low \textit{T}})}$  & 	$\makecell{\mu_{\rm eff} (\mu_{B})\\(\textnormal{ \rm low \textit{T}})}$  & $\makecell{T_{\rm N} (\rm K) \\ (\rm Ref.)}$ \\[2 ex ] \hline
		\makecell{YbMgGaO$_{4}$ \\ ($R\bar{3}m$) } & Yb-O-Yb & $-$4  & 2.8 &  \makecell{- \\ \cite{Li2015,Paddison2017}}    \\[2 ex] \hline
			\makecell{NaYbO$_{2}$ \\ ($R\bar{3}m$) } & Yb-O-Yb & $-$5.6  & 2.84 &  \makecell{-\\ \cite{PhysRevB.100.144432}}   \\[2 ex] \hline	
		
				\makecell{NaBaYb(BO$_{3}$)$_{2}$\\ ($R\bar{3}m$) } & Yb-O-B-O-Yb & $-$0.069  & 2.23 & \makecell{$0.41 $\\ \cite{PhysRevMaterials.3.094404}} \\[2 ex] \hline
					\makecell{K$_{3}$Yb(VO$_{4}$)$_{2}$\\ ($P6_{3}m$) } & Yb-O-V-O-Yb & $-$1  & 2.41 &  \makecell{-\\ \cite{PhysRevB.104.144411}}   \\[2 ex] \hline
					
					\makecell{NdTa$_{7}$O$_{19}$\\ ($P\bar{6}c2$) } & Nd-O-Ta-O-Nd & $-$0.46  & 1.9 &  \makecell{-\\ \cite{Arh2022}}\\[2 ex] \hline	
						\makecell{\textbf{Ba$_{6}$Yb$_{2}$Ti$_{4}$O$_{17}$}\\ \textbf{($P6_{3}/mmc$)} } & Yb-O-Ti-O-Yb & $-$0.49  & 2.5 &  \makecell{0.077 \\ Present}\\[2 ex] \hline			
	\end{tabular} 
	{\label{tableIN}}
\end{table} 
\subsection{Magnetic susceptibility}
In Figure~\ref{BYTO2} (a), we show the temperature dependence of magnetic susceptibility ($\chi(T)$)  in two magnetic fields in the temperature range 2 K $\leq$ \textit{T} $\leq$ 350 K. The magnetic susceptibility data do not show any signature of  long-range magnetic order at least above 2 K in BYTO. The absence of any marked difference  between zero-field cooled and field-cooled magnetic susceptibility data (see inset of Figure~\ref{BYTO2} (a)) suggests Yb$^{3+}$ spins are not frozen at least down to 2 K.  In order to calculate the value of effective magnetic moment ($\mu_{\rm eff}$) and Curie-Weiss temperature ($\theta_{\rm CW}$), the inverse $\chi(T)$ data (see Figure~\ref{BYTO2} (b)) were fitted by the Curie-Weiss (CW) law
$\chi= C$/($T-\theta_{CW}$).  Here, the Curie constant (\textit{C}) is associated with the effective moment through the formula $\mu_{\rm eff} = \sqrt{8C}  \mu_{B}$, and $\theta_{\rm CW}$ signifies the energy scale of the magnetic exchange interactions between Yb$^{3+}$ moments.  
 The high-temperature CW fit in the temperature range 150 K $\leq$ \textit{T} $\leq$ 350 K yields $\theta_{\rm CW}$ = $-$102 K and $\mu_{\rm eff}= 4.57 $ $\mu_{B}$. The large negative
Curie-Weiss temperature indicates the presence of elevated energy levels  of crystal electric field excitations. The obtained $\mu_{\rm eff}$ = 4.57 $\mu_{B}$ is
comparable to the effective moment of free Yb$^{3+}$ ions ($\mu_{\rm eff}^{\rm free}$ = 4.54 $\mu_{B}$). As previously mentioned, the Yb$^{3+}$ moments are located in an octahedral environment, and as a result, a strong crystal electric field naturally causes the eight-fold degenerate $J$ = 7/2 multiplet to split into four Kramers doublet states. \\
The preliminary indication of the presence of the ground-state Kramers doublet state can be observed in the temperature-dependent inverse susceptibility data. For instance, in BYTO, one can observe a deviation in the measured susceptibility data from the high-temperature Curie-Weiss fit (indicated by the red line) as depicted in Figure~\ref{BYTO2} (b), where this deviation becomes noticeable below 100 K. It is important to note that such deviations in transition metal-based magnets typically reflects the development of magnetic correlations between electronic spins, but in rare-earth magnets, it is typically due to presence of crystal electric field excitations.\\
To gain insights into the Kramers doublet ground state and the nature of magnetic interactions, it is necessary to perform a Curie-Weiss fit at low temperatures sufficiently  below the first excited crystal electric field level. It has been brought out in the literature that the  Curie-Weiss temperature for rare-earth-based magnets is strongly dependent on the temperature range for the Curie-Weiss fitting, owing to the influence of excited crystal electric field levels \cite{Arh2022,PhysRevB.108.054442}.\\  To estimate the nature of dominant
 magnetic interactions between Yb$^{3+}$ moments in the  ground-state Kramers doublet of BYTO, we performed fittings of the low-temperature inverse $\chi(T)$ data in various temperature ranges. In this analysis, the lower temperature limit was set at 4 K, while the upper temperature limit was systematically varied in 0.5 K increments up to 18 K (see inset of Figure~\ref{BYTO2} (b)) following a procedure as described in ref. \cite{Arh2022}.
 \begin{figure*}
 	\centering
 	\includegraphics[width=\textwidth]{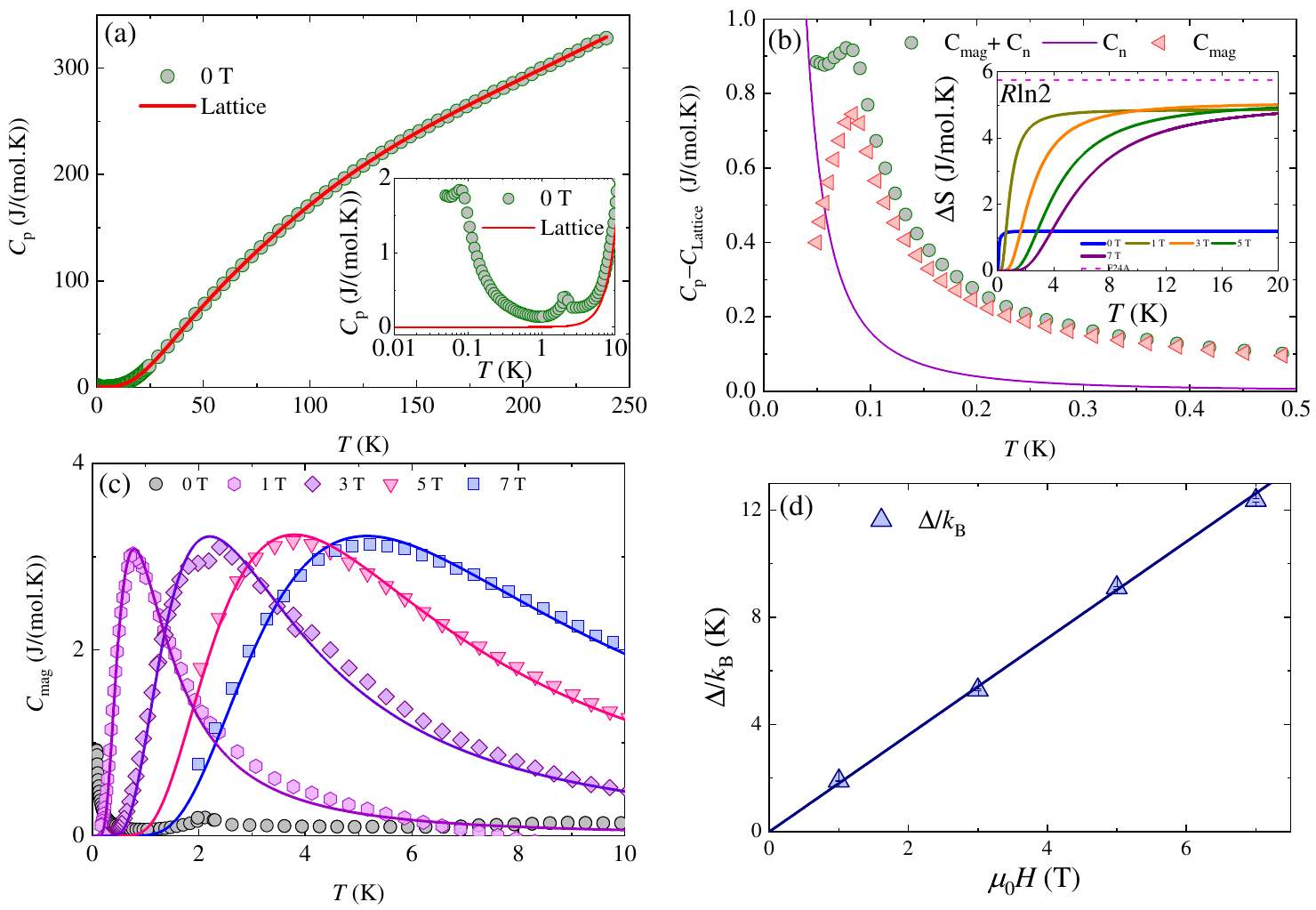}
 	\caption{(a) The temperature dependence of total specific heat ($C_{\rm p}$) of BYTO down to 49 mK in zero magnetic field. The solid line shows the Debye-Einstein model fit (see text) which represents the phonon specific heat. The inset depicts the appearance of two anomalies on decreasing the temperature below 3 K. (b) The temperature dependence of specific heat due to the sum of magnetic and nuclear contributions (circle) while the filled triangle represents the intrinsic  magnetic specific heat obtained after subtracting nuclear contributions (solid line). The inset shows the temperature dependence of entropy change up to 20 K in several magnetic fields. The dotted horizontal pink line is the expected entropy for $_{J_{\rm eff}}$ = 1/2 moment.
 		(c) The temperature dependence of the magnetic specific heat in several magnetic fields where the solid line represents two-level Schottky fit as described in the text. (d) The evolution of a field induced gap as a function of applied magnetic field where the solid line represents a linear fit. 
 	}{\label{BYTO3}}.
 \end{figure*}
 The estimated temperature independent fit parameter gives  $\mu_{\rm eff}$ = 2.51 $\mu_{B}$, which  is much smaller than $\mu_{\rm eff}$ = 4.54 $\mu_{B}$ resulting from the Hund's rule for a free  Yb$^{ 3+}$ (\textit{J} = 7/2) spin. This  indicates the formation of Kramers doublet ground state at low-temperatures and  $\theta_{\rm CW}$ = $-$0.49 K $ \pm $ 0.02 K which suggests the presence of a weak antiferromagnetic interaction between $J_{\rm eff}$ = 1/2 moments of Yb$^{3+}$ ions at low-temperature. Based on the effective magnetic moment ($\mu_{\rm eff}$) value of 2.51, the powder averaged Landé $g$ factor is found to be 2.89. In the mean-field approximation, the Curie-Weiss temperature ($\theta_{\rm CW}$) can be expressed as $\theta_{\rm CW} = (-zJS(S + 1))/(3k_{B})$, where $J$ represents the exchange interaction between the $J_{\rm eff}$ = 1/2 moment of Yb$^{3+}$ ions in the $ab$-plane, and $z$ denotes the number of nearest neighbors. In the case of BYTO, where $S$ = $J_{\rm eff}$ = 1/2 and $z$ = 6, the obtained value $J/k_{B}$ is 0.32 K \cite{Li2020}.
\\ Figure~\ref{BYTO2} (c) displays the magnetization curve at various temperatures.
		As corroborated by the specific heat data discussed below, BYTO behaves like a paramagnet above 1 K, therefore one can extract the powder-averaged Landé g-factor following the relation $M$/$M_{s}$ = $B_{1/2}$($y$), where $B_{J}(y) = [\frac{2J+1}{2J} \coth\left(\frac{y(2J+1)}{2J}\right)-\frac{1}{2J}\coth\left(\frac{y}{2J}\right)]$ represents the Brillouin function. Here, $M$ is the measured magnetization, $M_{s}$ (= $gJ\mu_{B}$) is the saturation magnetization, and $y = g\mu_{B}J \mu_{0}H/k_{B}T$, with $\mu_{B}$ denoting the Bohr magneton and $g$ representing the Lande g-factor. The solid lines in Figure~\ref{BYTO2} (c) correspond to the fitting of the Brillouin function, yielding an average value of g = 2.76, which is close to that estimated from the Curie-Weiss fit of low-temperature susceptibility data.
\subsection{Specific heat}
In order to unveil the magnetic ground state influenced by spin frustration and the correlation between Yb$^{3+}$ moments within the Kramers doublet ground  state with $J_{\rm eff}$ = 1/2 moments in BYTO, it is crucial to conduct experiments at mK temperatures owing to the weak exchange coupling between Yb$^{3+}$ moments. Hence, we performed specific heat measurements in a broad range of temperatures and magnetic fields.
 Figure~\ref{BYTO3} (a) depicts  the total specific heat data ($C_{\rm p}$) measured in a zero-magnetic field down to 49 mK. Below $T \leq$ 3 K, the specific heat data in zero magnetic field reveal two anomalies, one at 2.22 K and another at 77 mK, as depicted in the inset of Figure~\ref{BYTO3} (a). The 2.22 K anomaly arises from the long-range magnetic ordering due to the presence of unavoidable minor impurity phase of Yb$_{2}$O$_{3}$ \cite{PhysRevB.107.224416,PhysRevB.106.075132,PhysRevB.108.054442}. On the other hand the anomaly observed at 77 mK is attributed to the long-range magnetic order of Yb$^{3+}$ ions decorated on the triangular lattice. \\   
 In BYTO, the total specific heat can be represented as the combination of three components: the magnetic-specific heat ($C_{\rm mag.} (T)$) originating from magnetic Yb$^{3+}$ ions, the lattice-specific heat ($C_{\rm lat.} (T)$) attributed to the phonons, and the nuclear-specific heat resulting from the nuclear spins of Yb \cite{PhysRevB.93.100403}. In order to extract the entropy release resulting from the antiferromagnetic phase transition and the magnetic-specific heat, it is necessary to subtract the lattice and nuclear contributions to the total specific heat data. For this purpose, first, the lattice contribution was subtracted following the  Debye-Einstein model of lattice specific heat with one Debye term and three Einstein
 terms i.e., $
  C_{\rm lat.}(T)=C_{D}[9k_{B} \left(\frac{T}{\theta_{D}}\right)^{3}\int_{0}^{\theta_{D}/T}\frac{x^{4}e^{x}}{(e^{x}-1)^{2}}dx]
  +\sum_{i=1}^{3} C_{E_{i}}[3R\left(\frac{\theta_{i}}{T}\right)^{2}\frac{exp(\frac{\theta_{E_{i}}}{T})}{(exp(\frac{\theta_{E_{i}}}{T})-1)^{2}}], 
  $ 
 where $\theta_{D}$ is the Debye temperature, $\theta_{i}$ are  Einstein temperatures, and
 $R$ and $k_{B}$ are the molar gas constant and Boltzmann
 constant, respectively. 
 The good fit, as shown in  Figure \ref{BYTO3} (a),  suggests that the Debye-Einstein model reproduces well the lattice specific heat in this material \cite{kittel2005introduction}.  The best fit in the temperature range 30 K $\leq$ \textit{T} $\leq$ 160 K (see Figure~\ref{BYTO3} (a)) yielded the values $\theta_{D}$ = 179 K, $\theta_{E_{1}}$ = 260 K, $\theta_{E_{2}}$ =506 K,
 and $\theta_{E_{1}}$ = 1500 K \cite{gopal2012specific}.
   During the fitting procedure, the relative weight
   of acoustic  and 
   optical modes of vibration i.e., $C_{D}$ and $C_{E_{i}}$ were assigned at a fixed ratio of
$C_{D}$ : $C_{E_{1}}$ : $C_{E_{1}}$ : $C_{E_{2}}$ : $C_{E_{3}}$ = 1 : 2 : 3  : 8.5, which is consistent closely with the ratio of heavy atoms (Ba, Yb, Ti) and light atoms (O) in BYTO \cite{PhysRevB.93.100403,PhysRevB.106.104408}.
\\
The olive circle in Figure~\ref{BYTO3} (b) depicts the zero-field specific heat data obtained after subtracting the lattice contribution. The anomaly at $T_{N}$ = 77 mK indicates that BYTO undergoes an antiferromagnetic phase transition at low-temperature similar to the triangular lattice NaBaYb(BO$_{3}$)$_{2}$ \cite{PhysRevMaterials.3.094404}. The weak upturn in specific heat on decreasing temperature ($T$ $\leq$ 77 mK) can be attributed to a nuclear Schottky specific heat owing to the Yb nuclear spins akin to that observed in other Yb based magnets \cite{Sala_2023,PhysRevMaterials.3.094404}. To account for this contribution, we further subtracted the specific heat due to nuclear contribution following  $C_{n} \propto 1/T^{2}$ that is shown by the solid line in Figure~\ref{BYTO3} (b) \cite{PhysRevB.100.224417}. After subtracting the nuclear Schottky and lattice contributions, the resulting magnetic-specific heat data (represented by  triangles) is displayed in Figure~\ref{BYTO3} (b) as a function of temperature in zero-magnetic field. Below 1 K, there is a noticeable increase in magnetic-specific heat, followed by a sharp anomaly at 77 mK. This strongly suggests the presence of antiferromagnetic long-range order in BYTO. The occurrence of this anomaly at very low temperatures is in agreement with the presence of a weak exchange interaction between Yb$^{3+}$ moments.   \\ The inset in Figure~\ref{BYTO3} (b) shows the temperature-dependent magnetic entropy change, denoted as $\Delta S(T)$, which is calculated as the integral of $C_{\rm mag}(T)/T  dT$ in the several magnetic fields. In a zero magnetic field, below the lowest measured temperature, the magnetic specific heat was obtained using linear interpolation down to zero temperature.  Notably, it is observed that the entropy tends to attain a plateau at a value of 1.20 J/mol·K. This value corresponds to approximately 20\text{\%} of the expected entropy value of $R ln$ 2 (equal to 5.76 J/mol·K) for Yb$^{3+}$ spins with $J_{\rm eff}$ = 1/2 moments. The missing 80 \text{\%} entropy
can be due to the presence of short-range spin
correlations \cite{PhysRevB.104.L220403,PhysRevMaterials.3.094404,Khatua2021}. The overestimation of the lattice contribution to the missing entropy can be ignored as the experimental temperature range is sufficient  to adequately consider the impact of lattice contributions.  Furthermore, below the transition temperature $T_{N}$, the change in entropy amounts to approximately 5\text{\%} of $R \ln 2$. This observation implies the existence of substantial magnetic entropy below $T_{N}$, likely attributable to significant spin fluctuations with a small ordered moment \cite{Wu2019}. Furthermore, the spin fluctuation associated with short-range spin correlations above the transition temperature is also evident  from the $\mu$SR experiments as discussed in the next section. A common scenario of missing 80 \text{\%} entropy is also observed in recently reported  triangular lattice antiferromagnet Ba$_{3}$Yb(BO$_{3}$)$_{3}$ \cite{PhysRevB.104.L220403}. Nonetheless, the significant entropy retained at temperatures approaching absolute zero may arise from strong quantum fluctuations similar to the one observed in the long-range ordered triangular lattice antiferromagnet KBaGd(BO$_{3}$)$_{2}$ \cite{xiang2023dipolar,PhysRevB.107.125126}. Magnetic entropy tends to reach 5.76 J/molK in high magnetic fields, indicating the presence of $J_{\rm eff}$ = 1/2 moments.  
 \begin{figure}
	\centering
	\includegraphics[width=0.5\textwidth]{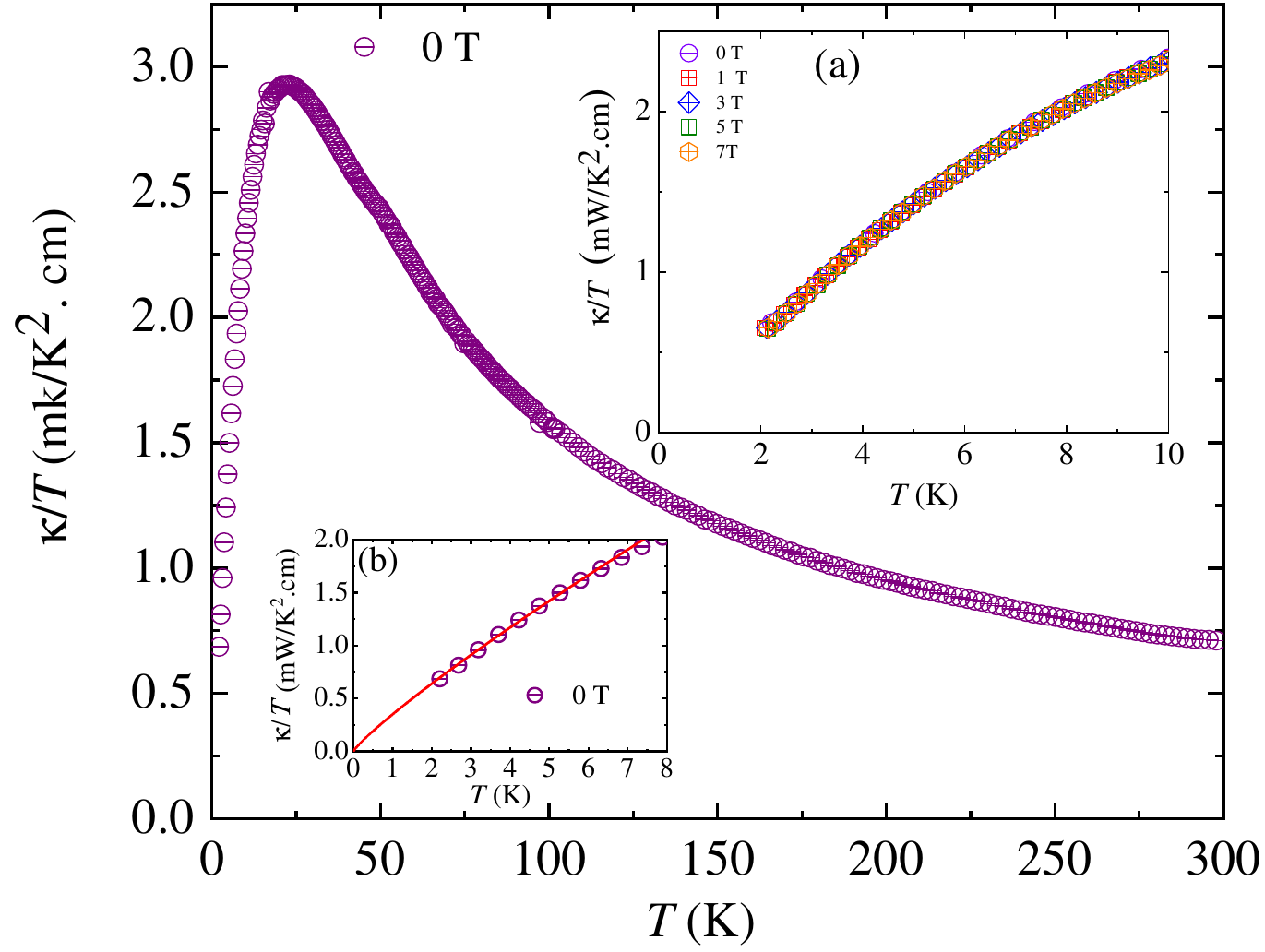}
	\caption{\hspace{-0.2cm} The temperature dependence of thermal conductivity divided by temperature ($\kappa/T$) of BYTO in zero magnetic field. The inset (a) shows $\kappa/T$ versus temperature in different magnetic fields up to 7 T. The  inset (b) displays $\kappa/T$ versus temperature in a zero-field where the red line represents the fitted phenomenological model described in the text.}
	\label{con}
\end{figure}
\\
In order to investigate the field  effect on the antiferromagnetically ordered ground state of BYTO, specific heat measurements were performed in different magnetic fields.  
After subtracting the lattice contributions, the obtained magnetic specific heat is shown in Figure~\ref{BYTO3} (c) for different magnetic fields.  It is observed that the anomaly at 77 mK disappears when a magnetic field of $\mu_{0}H$ = 1 T is applied.  This is a common feature observed in certain rare-earth magnets, including the compound under investigation, where strong magnetic field suppresses the exchange interactions. In such a scenario, the external magnetic field induces the Zeeman splitting of the lowest Kramers doublet state. Consequently, a Schottky-like broad peak emerges in the specific heat data as depicted in Figure~\ref{BYTO3} (c).
Furthermore, it is worth noting that as the strength of the applied magnetic field increases, the broad peak  shifts towards higher temperatures similar  to that observed in several other rare-earth magnets \cite{PhysRevMaterials.3.094404,PhysRevB.106.104404}.\\ 
  To deduce the gap induced by the Zeeman splitting of the lowest Kramers doublet due to the applied magnetic field, we employed the simplest two-level Schottky specific heat  model \cite{gopal2012specific} i.e.,
\begin{equation}
	C_{\rm sch}=f R \left(\frac{\Delta}{k_{B}T}\right)^2\frac{{\rm exp}(\Delta/k_{B}T)}{(1+{\rm exp}(\Delta/k_{B}T))^{2}},
	\label{scho}
\end{equation}
 where $\Delta$ is the Zeeman splitting of the ground state Kramers doublet of Yb$^{3+}$ ion, $k_{B}$ is the Boltzmann constant, \textit{R} is the universal  gas constant and \textit{f} measures the fraction of Yb$^{3+ }$ spins which contributes to the  splitting of the ground state doublet. In Figure~\ref{BYTO3} (c), the solid line illustrates that two-level Schottky fit effectively describes the magnetic-specific heat data acquired in several magnetic fields. Notably, the estimated Zeeman gap ($\Delta$) is observed to exhibit a linear variation with the external magnetic field, as depicted in Figure~\ref{BYTO3} (d).
  From a linear fit the value of the Landé $g$ factor, estimated to be $g$ = 2.66 $\pm$ 0.02, which is close to that determined from the magnetization data. Moreover, the Schottky fit yields a value of approximately $f$ $\sim$ 0.9 for non-zero magnetic fields. This indicates that nearly all of the Yb$^{3+}$ spins are involved in contributing to the Schottky specific heat in magnetic fields.
  \begin{figure}
  	\includegraphics[width=\columnwidth]{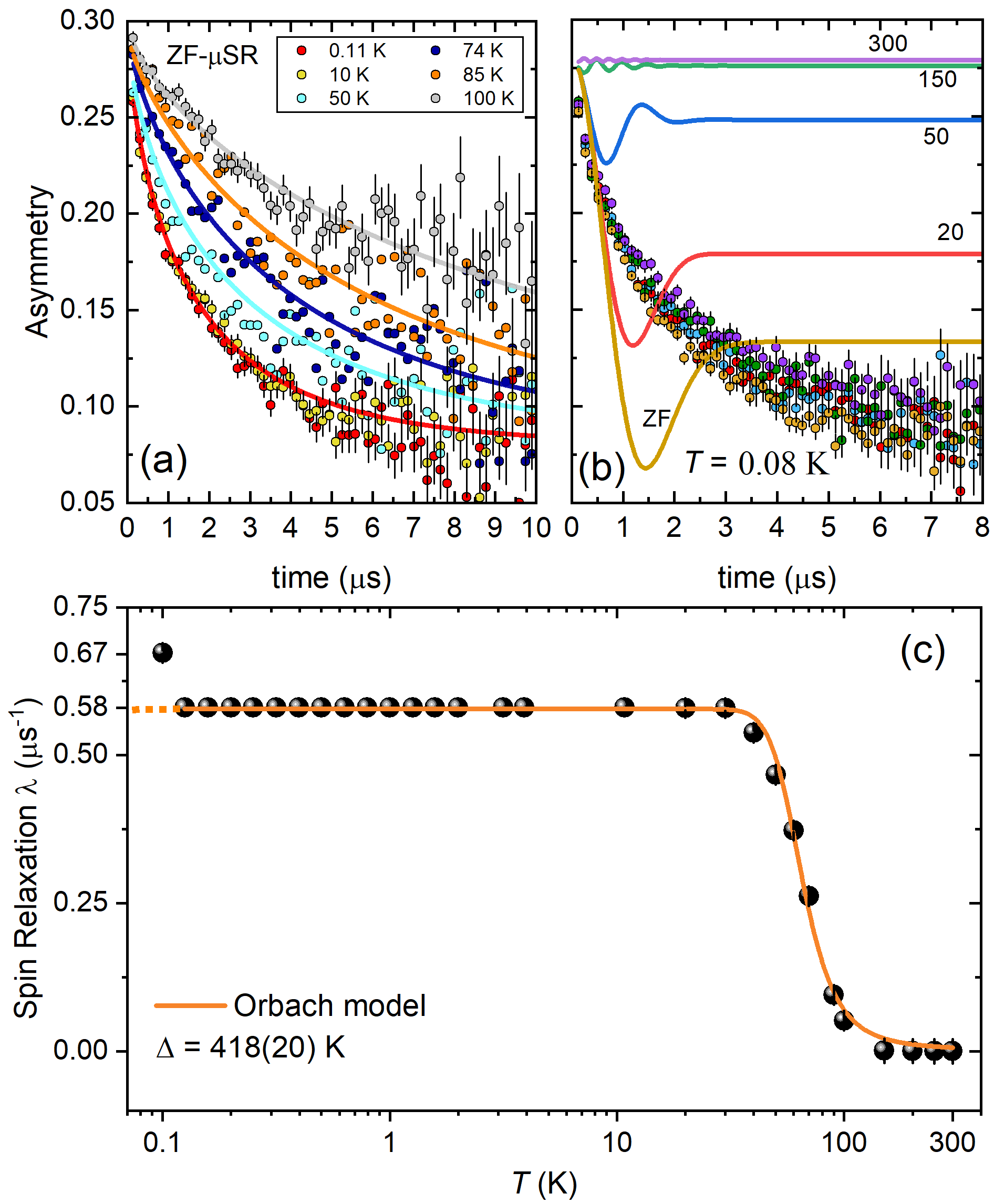}
  	\caption{\label{Fig_musr}
  		(a) Evolution in temperature of the zero-field asymmetry as a function of time for BYTO. Lines are fits with a stretched exponential. 
  		(b) Evolution of the asymmetry under applied longitudinal fields (circles from ZF to 300 Oe) and comparison with the behaviour of the static Kubo-Toyabe function for corresponding magnetic fields (lines, in Oe), highlighting the dynamical behaviour of Yb$^{3+}$ moments at 0.08~K.
  		(c) Temperature dependence of the muon spin relaxation rate $\lambda$ extracted from fits to the data shown in (a).  
  	}
  \end{figure}
\subsection{Thermal conductivity}
In correlated quantum materials, thermal conductivity  is a very sensitive probe to identify low-energy excitations which can carry heat or scatter heat carriers at low-temperature. Since the specific heat data are dominated by nuclear Schottky and lattice contributions, thermal conductivity is highly advantageous to probe the nature of ground state excitations, whether it is gapped or gapless. 
\begin{figure*}
	\centering
	\includegraphics[width=\textwidth]{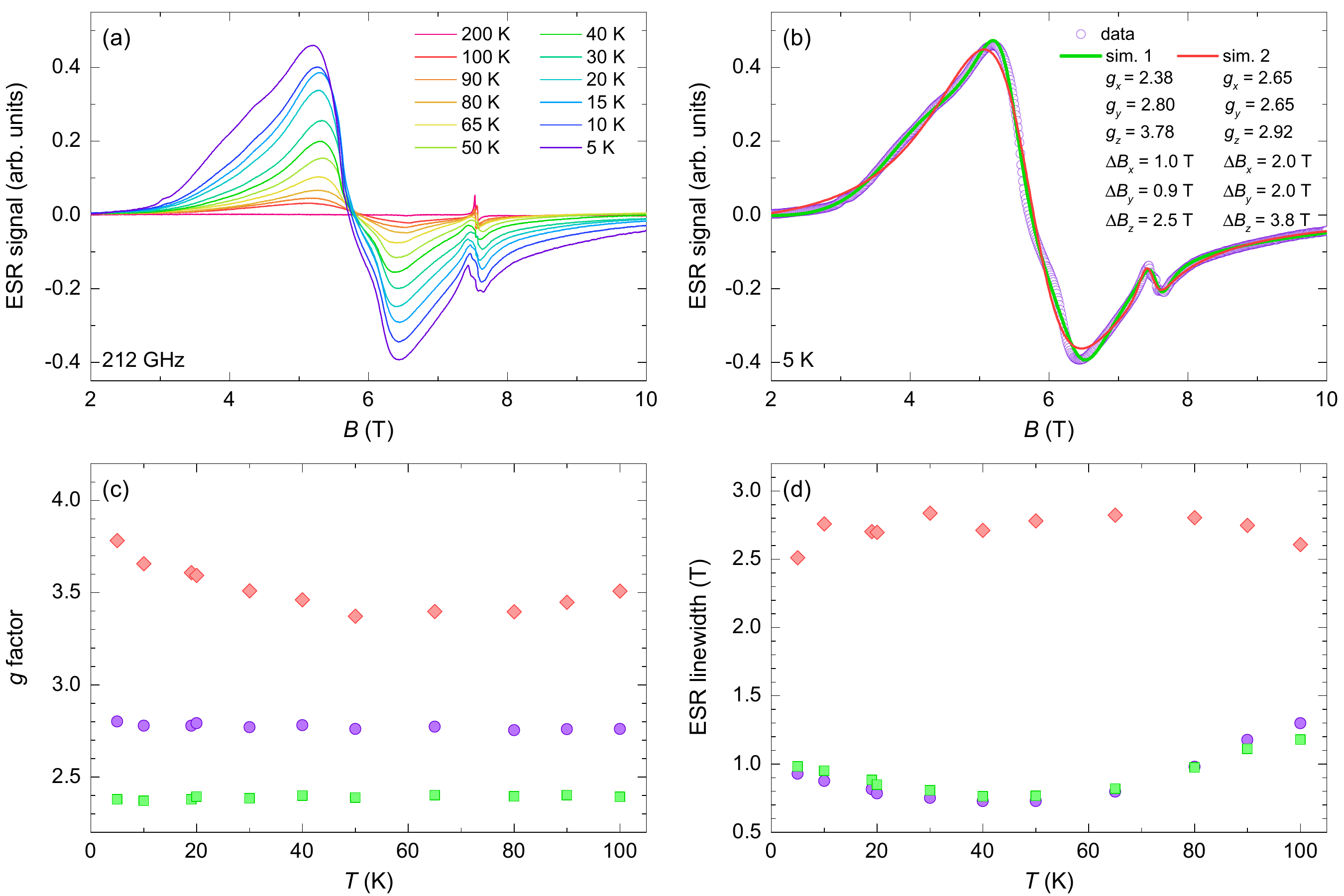}
	\caption{(a) The temperature evolution of the ESR spectra of BYTO. (b) Fit of the ESR spectrum at 5\,K (points) with a powder-averaged Lorentzian line shape with three independent $g$ factors (sim.~1) and two independent $g$ factors (sim.~2), the latter assuming uniaxial symmetry of the crystal lattice.
		(c) The temperature dependence of the three $g$ factor eigenvalues and (d) ESR line widths for sim~1.}
	\label{figESR}
\end{figure*}
 Figure~\ref{con} depicts the temperature dependence  thermal conductivity ($\kappa$) of a polycrystalline sample of BYTO. With decreasing temperature a well-defined broad peak appears around 20 K, which can be understood as the so-called phonon peak \cite{PhysRevX.9.041051,Boulanger2020,PhysRevResearch.4.L042035}. The appearance of such a peak is expected for typical insulating behavior of good quality polycrystalline samples. As depicted in inset (a) of Figure \ref{con}, there is no effect of magnetic field  on thermal conductivity, which implies that up to 2 K thermal conductivity is entirely dominated by phonons. Given that the interaction energy scale is approximately 0.30 K, magnetic excitations may potentially contribute to thermal conductivity at sub-Kelvin temperatures. For some insulating materials, thermal conductivity at low-temperature can be described as $\kappa/T$= $a$ + $b$ $T^{\alpha -1}$, where the first term represents the contribution of itinerant low-energy magnetic excitations and the second term is due to the contribution of phonons where the value of $\alpha$ typically  lies in the range 2 $\leq$ $\alpha$ $\leq$ 3 due to scattering of phonons from sample boundaries \cite{PhysRevB.77.134501,PhysRevB.67.174520}. As the measured temperature range is not sufficient to observe the contribution from magnetic excitations, the low temperature  $\kappa/T$ data were fitted  by  $\kappa/T$= $b$ $T^{\alpha -1}$  as shown in inset (b) of Figure~\ref{con}. 
The obtained value for $\alpha$ was found to be 1.87 $\pm$ 0.05, which is below the expected value \cite{ZHU2023100459,PhysRevLett.117.267202}. For triangular lattice antiferromagnets NaYbSe$_{2}$ and YbMgGaO$_{4}$, a similar values of $\alpha$ is also observed at lower temperature (\textit{T} $<<$ $\theta_{CW}$) \cite{ZHU2023100459,PhysRevLett.117.267202}.  
\subsection{Muon spin relaxation} 
In order to gain microscopic insights into the ground state spin dynamics, we performed highly sensitive muon spin relaxation measurements down to the base temperature of 80 mK. Figure~\ref{Fig_musr} depicts the zero field (ZF) and longitudinal field (LF) data, showing its evolution at several temperatures for BYTO.
As shown in Figure \ref{Fig_musr} (a), the ZF-$\mu$SR spectra exhibit no indications of either an oscillating component or a ``1/3" tail that reflects the lack of internal static magnetic fields down to 0.11 K. Nevertheless, the transition to magnetic ordering may not always be defined by oscillations in the zero field muon spin asymmetry, instead it can manifest as modifications in the asymmetry line shape \cite{PhysRevB.97.104409}. Neither this is observed in BYTO as all the ZF spectra are well fitted with a stretched exponential function, $A(t) = \exp[-(\lambda t)^{\beta}]+B$, with a moderate $\beta \simeq 0.8(1)$, and a background constant term $B$. The stretching parameter $\beta$ is found to be temperature independent and thus likely accounts for multiple relaxation rates due to a non-unique muon stopping site in the crystal structure. Upon cooling, the spin relaxation rate $\lambda$ reaches a plateau at a value of 0.58(3) $\mu$s$^{-1}$ below 10 K and down to 0.125 K. We noticed a slight increase of the muon spin relaxation
rate at 100 mK, perhaps indicative of a crossover towards
a more correlated regime, as pointed out
by the anomaly found at 77 mK in specific heat. 
The slowing down of spin fluctuations observed between 300 and 30 K  occurs in a typical temperature range of crystal electric field levels of Yb$^{3+}$~\cite{PhysRevB.92.134420}, to which the muon is sensitive through the Orbach process, with a muon spin relaxation rate modelled by~\cite{Arh2022,10.1007/3-540-30924-1_21}
\begin{equation}
\lambda =\left( C e^{-\Delta / k_B T}+\frac{1}{\lambda_0} \right)^{-1}
\end{equation}
\noindent with $\lambda_0 = 0.58(3)$~$\mu$s$^{-1}$ the relaxation rate when $T\rightarrow 0$, $C$ an amplitude parameter and $\Delta = 418(20)$~K the gap to the first excited crystal field level. The large value of  $\Delta$ ensures that only the Kramers ground-state doublet is occupied at low temperature and is well separated from the excited states and confirms the validity of the $J_{\rm eff} =1/2$ picture in BYTO. \\
In addition, a series of LF measurements were performed at 80 mK (Figure~\ref{Fig_musr} (b))  to gain insight into the dynamics of electronic magnetic moment.
 For large values of LF, in comparison with any static magnetic fields, the motion of muons spins is determined by LF only and the asymmetry becomes nearly constant. Consequently, any remaining depolarization of the muon spin is attributed to fluctuations due to electronic magnetic moments when LF overcomes local static fields, if any.  As shown in Figure \ref{Fig_musr} (b), even for large applied fields, the relaxation remains exponential. A crude comparison to a simple static Kubo-Toyabe relaxation form clearly fails to reproduce the data, and demonstrates the dynamical nature of the Yb$^{3+}$ moments at this temperature. A fluctuating regime is thus observed from 10~K down to 0.1~K with no slowing down below $\sim 4$~K, in  contrast to the related $J_{\rm eff}=1/2$ Yb triangular quantum spin liquid candidate YbMgGaO$_4$~\cite{PhysRevLett.117.097201}, but in agreement with the lower energy scale of the interaction in BYTO.
\subsection{Electron spin resonance}
 To study the nature of dominant magnetic interaction between Yb$^{3+}$ spins in BYTO, we performed ESR measurement on polycrystalline samples of BYTO.  
The ESR spectra contain a broad main component centered around $B_0=5.8$\,T, and a secondary narrow component centered at 7.54\,T (Figure~\,\ref{figESR} (a)). 
At 5\,K the intensity of the secondary component amounts to only 0.4\% of the main one and is therefore attributed to a minor impurity phase.
The main ESR line is composed of several pronounced features due to the polycrystalline nature of our sample.
A simulation of the powder-averaged spectrum works well with Lorentzian line shape and three independent $g$-factor components and line widths (simulation 1 in Figure~\,\ref{figESR} (b)), which is compatible with the distortion of the YbO$_6$ octahedra away from cubic symmetry.
At 5\,K, we find the $g$-factor eigenvalues $g_x = 2.38$, $g_y = 2.80$, and $g_z = 3.78$, and the corresponding line widths $\Delta B_x = 1.0$\,T, $\Delta B_y = 0.9$\,T, and $\Delta B_z = 2.5$\,T.
The fit is considerably worse if only two independent $g$ factors and line widths are considered (simulation 2 in Figure~\,\ref{figESR} (b)).
The latter model assuming axial symmetry around the crystallographic $c$ axis would apply for strongly exchange coupled Yb$^{3+}$ moments on the unit triangle \cite{gatteschi1990electron}.\\
The fact that the ESR spectrum does not obey the uniaxial symmetry of the underlying spin lattice limits the intra-trimer exchange interactions to $J\ll \mu_B \Delta g B_0/k_B\simeq 6$\,K, where $\mu_B$ and $k_B$ are the Bohr magneton and the Boltzman constant, respectively, and $\Delta g = g_z -g_x = 1.4$ denotes the spread of the $g$ factor at 5\,K.
The detected ESR linewidth is extremely broad and cannot be accounted for by simple dipolar interactions between Yb$^{3+}$ moments, which would result in ESR line width of the order $\mu_0\mu/(4\pi r^3)=11$\,mT, where $\mu_0$ is the vacuum permeability, $\mu=2.51\mu_B$ is the Yb$^{3+}$ magnetic moment in the ground state and $r = 5.91$\,\AA~is the nearest-neighbor distance.
Therefore, much larger magnetic anisotropy resulting from exchange interactions must also be present.
With increasing temperature the intensity of the ESR signal decreases profoundly (Figure~\,\ref{figESR} (a)), following a Curie-like dependence, which is in agreement with small exchange interactions.
The fit of the ESR spectrum becomes unreliable above 100\,K and the spectrum is completely lost at 200\,K.
We find that the ESR line widths (Figure~\,\ref{figESR} (d)) and $g$ factors (Figure~\,\ref{figESR} (c)), however, only slightly change with temperature, in agreement with the observation that the ground-state Kramers doublet is well separated from excited Kramers doublets.
Furthermore, phonon-related ESR broadening mechanisms which are all characterized by profound temperature dependence \cite{abragam2012electron} must be negligible in this temperature range.
\section{Discussion}
In rare-earth magnets, the electronic state of the rare-earth ion is mostly governed by the crystalline electric field at the rare-earth site and the number of electrons in the $4f$ shell.
Depending on such electronic state, crystalline anisotropy and favorable exchange path, rare-earth magnets can host distinct ground state properties \cite{PhysRevLett.120.207203}.  
In  rare-earth magnets with odd number of 4\textit{f } electrons, the crystal electric field usually splits the magnetic ground state of a free rare-earth ion into Kramers doublets, where 4\textit{f} ions acquire pseudospin- $J_{\rm eff}$ = 1/2 moments  in the ground-state doublet that is protected by time-reversal symmetry \cite{PhysRevX.11.021044}.
\\ In the present antiferromagnet BYTO (Yb$^{3+}$, 4$f^{13}$), crystalline electric field generated by nearby O$^{2-}$ ions  can  split the $2J+1$ = 8  degenerate ground state ($^{2}{F}_{7/2}$) into four Kramers doublets. The presence of these low-energy Kramers doublet states in BYTO has been indicated already through magnetic susceptibility measurements.   At low-temperatures, the magnetic susceptibility data follow a Curie-Weiss law  with effective moment $\mu_{\rm eff}$ = 2.51 $\mu_{B}$ that is much less than the effective moment of free Yb$^{3+}$ ions, which suggests the formation of Kramers doublet state with spin-orbit entangled $J_{\rm eff}$ = 1/2 moment of Yb$^{3+}$ ions. The estimated negative $\theta_{\rm CW}$ = $-$0.49 K from low-temperature susceptibility data indicates the presence of antiferromagnetic interaction though the strength of interaction is quite small which is typical of 4$f$ magnets \cite{PhysRevB.106.104408}.  
 The zero-field specific heat data clearly shows an anomaly at $T_{N}$ = 77 mK, which is tentatively assigned  to  antiferromagnetic long-range magnetic order in BYTO.  Nonetheless, the anomaly occurring at $T_{N}$ vanishes when subjected to a magnetic field with $\mu_{0}H \geq 1$ T, and instead, a broad maximum emerges at higher temperatures. This scenario is a common occurrence in rare-earth magnets, where the influence of Zeeman splitting resulting from an external magnetic field surpasses the weak exchange interactions between rare-earth moments, leading the system towards a field-polarized state \cite{PhysRevMaterials.3.094404,xiang2023dipolar}. 
As the magnetic field strength increases, the observed broad peak in specific heat for   $\mu_{0}H \geq 1$ T  broadens and shifts towards higher temperatures. This reflects the  Zeeman splitting of the lowest Kramers doublet state of BYTO, a phenomenon commonly observed in rare-earth magnets \cite{PhysRevB.108.054442,PhysRevB.106.104404}. The obtained 20 \text{\%} entropy in zero-magnetic field is less than the expected entropy ($Rln2$) for $J_{\rm eff}=$  1/2, suggesting there could be two origins of missing entropy; one is spin fluctuations in the ordered state and another is the presence of short-range spin correlation above the transition temperature due to moderate spin frustration defined by $f = |\theta_{\rm CW}|/T_{N}\approx 6$ \cite{PhysRevMaterials.3.094404,Khatua2021}. \\   
 In literature there are several reports on Yb-based triangular lattice antiferromagnets where the  exchange interaction between rare-earth moments is found to be the sum of dipolar and superexchange interactions \cite{xiang2023dipolar}. 
The intraplane Yb-Yb  distance in BYTO is almost double that of the intra-planar distance in YbMgGaO${4}$, which suggests the presence of weak magnetic dipole–dipole interaction approximately  0.017 K as estimated using   E$_{\rm dip}$$ \approx$ $\mu_{0}g_{\rm avg}^{2}\mu_{B}^{2}/4\pi a^{3}$ where $g_{\rm avg}$ is the powder average Land\'e $g$ factor and $a$ is the nearest-neighbor Yb-Yb distance in BYTO. The obtained dipolar interaction is only  5 \text{\%} of nearest-neighbor  exchange interaction as estimated from the Curie-Weiss temperature following mean-field approximation. This suggests the presence of dominant superexchange interaction between the $J_{\rm eff}$ = 1/2 moments of Yb$^{3+}$ ions. Our ESR results also indicate apart from dipolar interaction, the presence of finite exchange interactions in BYTO, which however is anisotropic.
 In BYTO, YbO$_{6}$ octahedra are isolated instead of corner and side sharing regular octahedra as observed in YbMgGaO$_{4}$.  Thus, in BYTO, the nearest neighbor intra-plane superexchange interaction can only be possible via the $ f-p-d-p-f$ (Yb-O-Ti-O-Yb) virtual  path, which is likely one of the reasons  of  weak antiferromagnetic interaction strength in BYTO. Whereas, in YbMgGaO$_{4}$, the nearest neighbor superexchange interaction is directly mediated by the oxygen ion via the $f-p-f$ (Yb-O-Yb) virtual electron hopping processes, as a result in YbMgGaO$_{4}$, antiferromagnetic interaction is a bit stronger ($\approx$ 2 K) \cite{PhysRevB.83.094411}. 
From crystallographic parameters, it is observed that the interplanar distance in BYTO is approximately 7.26 {\AA} and is slightly larger than the nearest-neighbor distance ($\sim$ 5.92 {\AA}) between rare-earth moments. Furthermore, in the triangular planes, the bond  distance of second nearest-neighbor of Yb$^{3+}$ ion  is roughly 10.23 {\AA} that is larger than the interplanar separation to make significant exchange coupling in BYTO. Therefore, it is most likely that in addition to intraplane nearest-neighbor exchange interactions, the presence of non-negligible interplane exchange interactions through the Ti$_{2}$O$_{9}$ dimer lead to long-range magnetic order in this triangular lattice antiferromagnet. An antiferromagnetic phase transition due to interlayer interaction  is also observed in  triangular lattice antiferromagnet Ba$_{3}$CoSb$_{2}$O$_{9}$ (Co$^{2+}$; $J_{\rm eff} = 1/2$), which has similar  crystallographic symmetry, intralayer, interlayer exchange paths, and interlayer distance with that of BYTO \cite{PhysRevLett.108.057205,Kamiya2018}. \\
  $\mu$SR experiments   reveal a fluctuating state of Yb$^{3+}$ moments  in the temperature range  0.1 K $\leq$ $T$ $\leq$ 10 K and  do not show any signature of long-range magnetic  order down to 80 mK.
  The broadening of ESR spectrum at 5 K indicates the presence of  anisotropy in the exchange interaction between the Yb$^{3+}$ spins, otherwise one would observe narrow ESR spectrum for isotropic exchange interaction \cite{RevModPhys.25.269}.  We also found large  $g$- factor anisotropy, which is compatible with the distortion of the YbO$_{6}$ octahedra from cubic symmetry. The temperature independent $g$ factor in the broad temperature range implies the well separated  lowest Kramers doublet state that is consistent with $\mu$SR results. The estimated $g$ value from ESR spectrum at 5 K leads to powder-average value  $g$ = $\sqrt{(g_{x}^{2}+g_{y}^{2}+g_{z}^{2})/3}$ = 3.04, which is close to that obtained from magnetization data. Theoretically, it is predicted  that apart from interlayer exchange interaction, the nearest-neighbor exchange interaction with either easy-plane or easy-axis anisotropy can lead to long-range ordered state in rare-earth based triangular lattice antiferromagnet \cite{PhysRevLett.120.207203}. The role of anisotropic exchange interaction in stabilizing long-range order state is not yet clear from  the present study and will be a subject of detailed
future investigations.
 \section{Conclusion}
 We presented the  synthesis, crystal structure, thermodynamic, muon spin relaxation, and electron spin resonance results of a novel triangular lattice antiferromagnet Ba$_{6}$Yb$_{2}$Ti$_{4}$O$_{17}$, which crystallizes in the highly symmetric crystal structure with space group $P6_{3}/mmc$ without any detectable anti-site disorder between atomic sites. The present compound constitutes close to structurally perfect two dimensional triangular layers of Yb$^{3+}$ ion  perpendicular to the crystallographic  \textit{c}-axis. The magnetic susceptibility and electron spin resonance measurements suggest the presence  of weak antiferromagnetic interaction between $J_{\rm eff}= 1/2 $ moment of  Yb$^{3+}$ ions. The anomaly observed at $T_{N}$ = 77 mK  in zero-field specific heat data is attributed to the presence of long-range magnetic order   which vanishes in magnetic field $\mu_{0}H$ $\geq$ 1 T. Furthermore,  the specific heat data suggest the presence of a field induced gap due to Zeeman splitting of the Kramers doublet ground state in weak magnetic field. 
$\mu$SR experiments reveal a fluctuating  regime in 0.1 K $\leq$ $T$ $\leq$ 10 K ascribed to the depopulation of crystal electrical field levels, in agreement with a well seprated Kramers doublet,  and the enhancement of muon spin relaxation rate below 100 mK might be  associated with a more  correlated  regime of Yb$^{3+}$ moments that is consistent with the presence of an anomaly in specific heat. Our ESR results further confirm that the lowest Kramers doublet is
well separated from the excited doublet state and point to  the
presence of anisotropic exchange interaction between
Yb$^{3+}$ spins in the present antiferromagnet. Despite the moderate spin frustration, a combination of exchange anisotropy and interlayer magnetic interaction are most likely to stabilize long-range magnetically ordered state in Ba$_{6}$Yb$_{2}$Ti$_{4}$O$_{17}$.
Future experiments may shed  detailed insights into the microscopic Hamiltonian, precise nature of anisotropy and spin correlations in this class of spin-orbit driven $J_{\rm eff}$ = 1/2 triangular lattice  antiferromagnet.\\
\begin{center}
 \textbf{ACKNOWLEDGMENTS}	
\end{center}
 P.K. acknowledges funding by the Science and Engineering Research Board and Department of Science and
 Technology, India through research grants. A.Z. acknowledges the financial support of the Slovenian Research and Innovation Agency
 through Program No. P1-0125 and Projects No. J1-50008 and
 No. BI-US/22-24-065.
 	E.K. acknowledges financial support from the labex
 PALM for the QuantumPyroMan project (ANR-10-
 LABX-0039-PALM). A.M.S. thanks the URC/FRC of UJ and the SANational Research Foundation (93549) for financial support.  The National High Magnetic Field Laboratoryis supported by National Science Foundation throughNSF/DMR-1644779 and the State of Florida.
\bibliographystyle{apsrev4-1}
\bibliography{BYTO}
\end{document}